\documentclass[]{aastex701}

\usepackage{graphicx} 
\usepackage{amsmath}

\begin{document}

\title{Clustering SPH Debris into N‑body Fragments: A Collisional Code for Planet Formation}

\author[]{Samuele Crespi}
\affiliation{Center for Astrophysics and Space Science (CASS), New York University Abu Dhabi, PO Box 129188, Abu Dhabi, UAE}
 \email{}

\author[]{Mohamad Ali-Dib} 
\affiliation{Center for Astrophysics and Space Science (CASS), New York University Abu Dhabi, PO Box 129188, Abu Dhabi, UAE}
\email[show]{malidib@nyu.edu}

\author[]{Ian Dobbs-Dixon}
\affiliation{Center for Astrophysics and Space Science (CASS), New York University Abu Dhabi, PO Box 129188, Abu Dhabi, UAE}
\email{}

\begin{abstract}
Giant impacts among planetary embryos generate long–lived debris that feeds back on late-stage terrestrial planet growth, yet most $N$-body models either assume perfect merging or treat fragments in ad hoc ways. We present \texttt{SHARD}, a collision–resolution framework that couples a hybrid  integrator (\textsc{REBOUND}/\texttt{mercurius}) to a six‑dimensional catalog of smoothed‑particle hydrodynamics (SPH) outcomes spanning impact speed, geometry, total mass (up to $2 M_{\oplus}$), mass ratio, and target/projectile water fractions. For each detected impact we multi‑linearly interpolate  to return the two largest remnants with self‑consistent kinematics and volatile budgets, and we reconstruct the unresolved fragment population by aggregating nearby SPH debris snapshots and compressing them with mass‑weighted $k$‑means in velocity space into a tractable number of fragments above a tunable minimum mass. Exact conservation of total mass and water mass is enforced across survivors and debris, with immediate, energy‑based reaccretion checks performed within the timestep. Debris interpolation is constrained to the tabulated SPH grid (no extrapolation), and special handling of hit‑and‑run and catastrophic regimes is included.
We benchmark our code against \texttt{SyMBA} in a Mercury‑formation experiment and find broad qualitative agreement, though our final embryos distribution is dynamically hotter and more top-heavy in mass.
{The benchmark outcome suggests that debris clustering is dynamically consequential: fragment resolution controls reaccretion, damping, and volatile retention.}
This SPH‑anchored debris treatment provides a drop‑in, compositionally aware alternative to perfect merging, enabling late‑stage accretion studies that retains fragment feedback without saturating the integrator.

\end{abstract}


\section{Introduction}
\label{sec:intro}

The final assembly of terrestrial planets proceeds through a sequence of giant impacts among Moon-- to Mars--sized planetary embryos embedded in a dynamically excited inner disk  \citep{Chambers2001,OBrien2006}. As collisions accumulate, a non-negligible fraction of the solid inventory is converted into a spectrum of fragments spanning dust to lunar sizes. These fragments are neither instantaneously re-accreted nor negligibly small; their long-lived dynamical and collisional processing can reshape the growth history and final architectures of rocky worlds  \citep{AgnorAsphaug2004,LeinhardtStewart2012}. In particular, imperfect merging outcomes such as hit-and-run collisions are expected to be common— \cite{KokuboGenda2010} find that nearly half of embryo--embryo impacts result in hit-and-run events—implying that perfect accretion is an over-simplification of late-stage growth. A central challenge for predictive models is therefore the physically faithful treatment of collisions and post-collisional debris within $N$-body simulations that, by necessity, cannot resolve all size scales  \citep{KobayashiDauphas2013}.

Modern $N$-body studies commonly adopt one of four classes of collision solvers. The historically prevalent \emph{Perfect Merging} scheme treats every impact as a perfectly inelastic merger, an assumption now known to shorten formation timescales and overestimate final planet masses and water budgets  \citep{Chambers2001,AgnorAsphaug2004}. More realistic approaches allow for fragment production. \emph{Unresolved Debris} methods, for example, preserve the two largest collision remnants and deposit the remaining mass into one or more tracers (massless or low-mass particles) to represent a cloud of debris  \citep{LevisonThommesDuncan2012,Chambers2013}. \emph{Same-Mass Debris}  prescriptions distribute the fragmented mass into numerous equal-mass super-particles guided by scaling laws for collisional outcomes  \citep{Chambers2013}, ensuring a computationally manageable number of fragments. In contrast, \emph{Fragmentation without Debris}  models incorporate the results of high-resolution impact calculations to determine the fate of the largest remnants (often the largest and second-largest bodies) while excising the myriad smaller fragments to avoid saturating the integrator  \citep{MarcusStewartSasselovHernquist2009,StewartLeinhardt2012}. Collectively, these more sophisticated approaches demonstrate that collisional fragmentation can roughly double the accretion timescale and reduce final planet masses relative to perfect merging scenarios  \citep{Chambers2013,WalshLevison2019,BurgerBazsoSchafer2020}. Nevertheless, a homogeneous inter-comparison of collision treatments and, crucially, a debris model calibrated to first-principles impact physics across the full fragment size spectrum, have remained pressing needs in the quest to faithfully simulate the assembly of terrestrial planets.


In this paper we develop and implement \texttt{SHARD} (SPH-Hydrodynamic Accretion and Remnant Debris): a collision–resolution scheme that maps $N$-body impacts onto smoothed-particle hydrodynamics catalogs and statistically reconstructs the unresolved fragment population through mass-weighted k-means clustering to provide a framework that explores the dynamical and compositional consequences of debris production during late-stage planetary accretion.
{Our work builds upon a catalog {\citep{Schafer2016,Schafer2020}}} of smoothed particle hydrodynamics (SPH) collision outcomes spanning the dominant region of parameter space for embryo–embryo impacts \citep{Benz1994,Monaghan1985, Schafer2016}. This enables a statistical characterization of the mass, size, and orbital distributions of post-impact debris as functions of total mass, mass ratio, velocity, and geometry, and anchors debris prescriptions directly to hydrodynamic physics rather than purely empirical extrapolation \citep{Leinhardt2012}.

\section{Methods}
\label{sec:methodsdata}
\subsection{Overview and numerical units}

We model late-stage planetary assembly with a hybrid $N$-body integrator that resolves pairwise impacts using outcomes calibrated to smoothed-particle hydrodynamics (SPH) experiments, and that replaces unresolved collision debris by a small number of self-consistent fragments obtained via mass-weighted clustering in velocity space. The implementation is built on top of \textsc{REBOUND} \citep{rebound} with the hybrid \texttt{mercurius} integrator \citep{mercurius}, and leverages \textsc{REBOUNDX} \citep{reboundx} to carry passive per-particle parameters (notably the bulk water fraction, \texttt{wf}). All quantities are expressed in \textsc{REBOUND} units with $G=1$, distance in AU, mass in $M_{\odot}$, and time in yr$/2\pi$.\footnote{In these units, the Earth's mass is $M_{\oplus}\equiv\mathrm{MEAR}=3.003\times10^{-6}\,M_{\odot}$.}

The code tracks bodies gravitationally, and these are split into active and test particles via the \texttt{N\_active} parameter in \textsc{REBOUND}. Collisions are identified with the \texttt{direct} geometry, and close encounters are handled by switching locally to the high-accuracy \texttt{IAS15} integrator. The collision resolution module queries an SPH outcome catalogue to (i) return the two largest post-impact bodies via multi-linear interpolation across the SPH grid (Sect.~\ref{sec:sph-table}) and (ii) reconstruct the velocity distribution of small fragments by aggregating neighbouring SPH debris snapshots and clustering them into a prescribed number of fragments (Sect.~\ref{sec:debris-catalogue}--\ref{sec:kmeans}). Throughout, we enforce mass and water-mass conservation as described below. 

\subsection{Integrator configuration and outputs}
\label{sec:integrator}

By default, we adopt \texttt{mercurius} with Hill switching factor $h_{\mathrm{fac}}=5$ and global time step $\Delta t=10^{-2}\times 2\pi$ (i.e. $0.01$\,yr); during close encounters the minimum \texttt{IAS15} step is restricted to $10^{-4}\,\Delta t$. Collisions are handled by a custom resolver \texttt{collision\_solver}. Every checkpoint we write system snapshots, a human-readable events log, and a machine-readable collisions table. Snapshots and events include each body's mass, orbital elements, water fraction, Cartesian state, and a \emph{code} label to track identity (Sun, gas giants, embryos, planetesimals, debris).

To emulate an open system, by default,  we remove bodies according to: (i) unbound orbits ($e\ge 1$), (ii) semimajor axis $a>15$~AU, (iii) orbits too close to the Sun ($a<0.1$~AU or perihelion $q<0.01$~AU). Removed particles are recorded.

\subsection{Collision parameterisation}
\label{sec:coll-params}

Let the pre-impact bodies have masses $m_1$ and $m_2$, positions $\boldsymbol{x}_1,\boldsymbol{x}_2$, and velocities $\boldsymbol{v}_1,\boldsymbol{v}_2$. We define total mass $M_{\rm tot}=m_1+m_2$, mass ratio proxy
\begin{equation}
\gamma \;=\; \min\!\left(\frac{m_1}{m_2}, \frac{m_2}{m_1}\right)\in (0,1],
\end{equation}
and the relative position and velocity, $\Delta\boldsymbol{x}=\boldsymbol{x}_1-\boldsymbol{x}_2$ and $\Delta\boldsymbol{v}=\boldsymbol{v}_1-\boldsymbol{v}_2$. The mutual escape speed at contact is $v_{\rm esc}=\sqrt{2M_{\rm tot}/\|\Delta\boldsymbol{x}\|}$. We work with the dimensionless impact speed $v_0 = \|\Delta\boldsymbol{v}\|/v_{\rm esc}$ and impact angle
\begin{equation}
\alpha \;=\; \arccos\!\left(\frac{|\Delta\boldsymbol{x}\cdot \Delta\boldsymbol{v}|}{\|\Delta\boldsymbol{x}\|\,\|\Delta\boldsymbol{v}\|}\right) \in [0,\pi/2],
\end{equation}
so that head-on impacts have $\alpha\simeq 0$ and grazing impacts have $\alpha\simeq \pi/2$. Each impactor carries a bulk water mass fraction $w_{\rm f}$ (in \%), propagated as a passive scalar (\texttt{wf}). The six-tuple $(v_0,\alpha,M_{\rm tot},\gamma,w_{\rm f,t},w_{\rm f,p})$ provides the query point for the SPH outcome library (Sect.~\ref{sec:sph-table}), where subscripts $t$ and $p$ refer to target and projectile respectively.

\subsection{SPH outcome library and encoding}
\label{sec:sph-table}
{
The SPH dataset consists of a grid of simulations spanning discrete values of the six control parameters above.
We used a {dataset of homogeneous SPH simulations} from the works of 
\cite{Schafer2016,Schafer2020}. It consists of 880 low-resolution SPH simulations (about 20000 particles per scenario) of two colliding protoplanets.} {Database and parameter ranges are described in detail in Appendix \ref{sec:data}.}

Each simulation is assigned a base-$b$ code \texttt{v0/alpha/mtot/gamma/wft/wfp}, and stores (i) the two largest post-impact bodies (CoM-centric positions and velocities in spherical coordinates, mass fraction, and water mass fraction \emph{relative to the total pre-impact water mass}) and (ii) a debris catalogue file with all fragments as Cartesian $(\boldsymbol{x},\boldsymbol{v})$, {i.e. the complete post-impact fragment list beyond the two largest remnants down to the smallest entries present in the SPH output (no additional ``resolved'' cut is applied at this stage)}, per-particle water fraction, and a weight proportional to the particle mass fraction. {For catalogue entries we further store, per SPH collision, a flag indicating “perfect merging’’ (PM; second body absent) and an internal flag for catastrophic disruption, denoted ``crashed'' in the code. Here ``crashed'' means \texttt{Nbig}$=-1$: the remnant/debris payload for that SPH catalogue entry is treated as unavailable for interpolation; it does not refer to a failed $N$-body integration. The code reads the global \texttt{SPH.table} to assemble a list of collisions and loads per-collision debris files from \texttt{SPHDebris\_catalogue/}.}

\paragraph{Interpolation of the two largest remnants.}
Given a point \\
$\boldsymbol{\theta}=(v_0,\alpha,M_{\rm tot},\gamma,w_{\rm f,t},w_{\rm f,p})$, {we locate the two neighbouring nodes per dimension and compute a sequence of binary interpolations across the $2^6$-corner hypercube. The circular interpolation below is not applied to all six control parameters; rather, it is used for the angular components of the remnant positions and velocities stored in \texttt{SPH.table} as spherical coordinates, $[r,\theta,\phi]$ and $[v,\theta_v,\phi_v]$. For these angles we perform \emph{circular} interpolation using}
\begin{equation}
\mathrm{interp}_{\angle}(a,b;\,d)\;=\;\arctan2\!\Big[(1-d)\sin a+d\sin b,\;\,(1-d)\cos a+d\cos b\Big],
\end{equation}
{This avoids artificial discontinuities at the angular branch cut. We adopt logarithmic interpolation along the $M_{\rm tot}$ axis.} Special cases are handled to ensure physicality: if either corner indicates PM (second body mass $=-1$ in the table), we interpolate to or from the PM state consistently; we clip negative masses, renormalise the sum of the two mass fractions to unity if necessary, and clip negative/greater-than-unity water fractions (with a final renormalisation step if $w_1+w_2>1$). {The routine returns the two largest bodies as $(\boldsymbol{x}_i^{\rm sph},\boldsymbol{v}_i^{\rm sph},\,m_i/M_{\rm tot},\,W_{i}/W_{\rm tot})$, where $\boldsymbol{x}_i^{\rm sph}$ and $\boldsymbol{v}_i^{\rm sph}$ are the position and velocity \emph{3-vectors} of remnant $i$ in the SPH collision centre-of-mass (CoM) frame. In \texttt{SPH.table} they are stored as spherical components $[r,\theta,\phi]$ and $[v,\theta_v,\phi_v]$ and are converted to Cartesian vectors before being mapped into the global simulation frame (Sect.~\ref{sec:frames}).}

\paragraph{Counting macroscopic survivors.}
{ We enforce a minimum resolved fragment mass $M_{\rm FM}$ (default $5.5\times10^{-4} M_{\oplus}$) that limits the number of macroscopic survivors. Let the interpolated largest-body mass fractions be $\mu_1\ge \mu_2$ ($\mu_1$=$M_1$/$M_{tot}$). We classify the outcome as (i) catastrophic if $\mu_1<0.1 (1+\gamma)^{-1}$, (ii) \emph{projectile accretion/destruction} if $\mu_2<0.1 \gamma(1+\gamma)^{-1}$, and (iii) hit-and-run otherwise. Survivors with $m_i<M_{\rm FM}$ are dropped from the list of resolved bodies, their mass being added to the debris reservoir. Water is then rebalanced to enforce $0\le W_i\le m_i$ and $W_{\rm fr}\le M_{\rm fr}$. The remaining mass fraction, $M_{\rm fr}/M_{\rm tot}$ and $W_{\rm fr}/W_{\rm tot}$, is reserved for fragments (Sect.~\ref{sec:debris-catalogue}).

In our implementation, $M_{\rm FM}$ also fixes the debris multiplicity through $N_{\rm fr}=\lfloor M_{\rm fr}/M_{\rm FM}\rfloor$, and, together with the activation threshold, defines the effective mass-space resolution of the post-impact cascade. Because the clustered debris are then renormalized back to the same SPH-interpolated $M_{\rm fr}$ and $W_{\rm fr}$, changing $M_{\rm FM}$ {does not alter the total collisional mass or volatile budget, but rather how that fixed budget is discretized dynamically.} {Larger $M_{\rm FM}$ therefore concentrates the same debris mass into fewer, more massive tracers, increasing discreteness and making the local energy-based reaccretion criterion easier to satisfy for individual clumps. This shortens the lifetime of the low-mass debris reservoir and tends to produce fewer leftover small bodies together with a more top-heavy and dynamically hotter terrestrial-planet outcome}. Smaller $M_{\rm FM}$, conversely, distributes the same debris budget over more numerous lower-mass bodies and should {push the evolution closer to that of Scora et al. 2024. Fragment-resolution choice may therefore contribute to residual differences between simulations even in the absence of different underlying collision physics.}}

\paragraph{Water conservation and clipping for survivors}
\label{alg:water}
For each survivor $i$, clip $W_i \leftarrow \min\!\bigl[\,W_i,\; m_i / w_{\rm tot}\,\bigr]$, where $w_{\rm tot}\equiv W_{\rm tot}/M_{\rm tot}$. With $M_{\rm fr}=1-\sum_i (m_i/M_{\rm tot})$ and $W_{\rm fr}=1-\sum_i (W_i/W_{\rm tot})$, set $W_i\leftarrow 0$ for all $i$ if $W_{\rm tot}=0$; otherwise, if $W_{\rm fr}> M_{\rm fr}/w_{\rm tot}$, scale all $W_i$ by the common factor $(1 - M_{\rm fr}/w_{\rm tot})/(1 - W_{\rm fr})$.

\subsection{Debris catalogue interpolation}
\label{sec:debris-catalogue}

{For the small-fragment phase space we combine the debris from the $2^6$ corners of the query hypercube but \emph{without} extrapolation beyond the tabulated grid. For a corner $c=(c_1,\ldots,c_6)$, with $c_k=0$ or $1$ denoting the lower or upper node along dimension $k$, we define $\lambda_k(c_k)=1-d_k$ for $c_k=0$ and $\lambda_k(c_k)=d_k$ for $c_k=1$, where $d_k$ is the one-dimensional interpolation coordinate. The corresponding corner weight is $\Lambda_c=\prod_{k=1}^{6}\lambda_k(c_k)$.} {For each contributing SPH collision we retain all catalogue entries beyond the two largest remnants (i.e. all rows after the first \texttt{Nbig}), so the catalogue includes the full SPH fragment list down to the smallest entries present in the SPH output rather than only ``resolved'' tertiary bodies.} {Corners flagged as ``crashed'' are likewise excluded from the debris phase-space sampling because no usable debris catalogue is available. The debris phase-space is therefore assembled from the remaining non-crashed contributing corners within the tabulated grid and is subsequently renormalized to the residual debris mass and water budgets, $M_{\rm fr}$ and $W_{\rm fr}$, computed after the remnant interpolation and conservation step.} Each debris entry contributes
\[
\big(\boldsymbol{x},\,\boldsymbol{v},\, \tilde{m},\, w_{\rm f}\big) \quad\text{with}\quad \tilde{m}=\Lambda_c\times \frac{m_{\rm SPH}}{M_{\rm tot}},
\]
where $m_{\rm SPH}$ is the {particle mass reported in the catalogue, and $M_{\rm tot}$ is the total mass involved in the collision. Because debris are treated as a weighted phase-space sample, adjacent grid points may contain widely different numbers of catalogue entries; we therefore do not attempt any one-to-one matching of fragments between SPH runs and simply concatenate the variable-length lists, with $\tilde{m}$ encoding each corner’s relative contribution.} The union of these lists forms a weighted sample that approximates the local debris distribution under mild parameter changes, and is written to a diagnostic ``\texttt{.debris}'' file for reproducibility.

\subsection{Mass-weighted clustering in velocity space}
\label{sec:kmeans}

To reduce the debris to a tractable number of bodies, we cluster the interpolated particles in \emph{velocity} space. The number of desired clusters is
\begin{equation}
N_{\rm fr} \;=\; \left\lfloor \frac{M_{\rm fr}}{M_{\rm FM}} \right\rfloor,
\end{equation}
with $M_{\rm fr}=M_{\rm tot}\times$ (interpolated debris mass fraction) and $M_{\rm FM}$ the minimum resolved fragment mass in Earth masses (Sect.~\ref{sec:sph-table}). We then run \texttt{KMeans} on the three-dimensional set $\{\boldsymbol{v}_j\}$ with \emph{sample weights} $w_j=\tilde{m}_j$ (proportional to the particle masses). For each cluster $c$ we compute mass-weighted averages of position and velocity,
\begin{equation}
\bar{\boldsymbol{x}}_c =
\frac{\sum_{j\in c} w_j \bar{\boldsymbol{x}}_j}{\sum_{j\in c} w_j},
\qquad
\bar{\boldsymbol{v}}_c =
\frac{\sum_{j\in c} w_j \bar{\boldsymbol{v}}_j}{\sum_{j\in c} w_j}.
\end{equation}
{ and the cluster water fraction as the mass-weighted mean of $w_{{\rm f},j}$. Because SPH-derived water partitioning for small remnants is resolution-limited, these per-cluster water fractions are primarily used to conserve the debris water mass budget; we therefore recommend interpreting the debris composition chiefly through the mass-weighted mean water fraction of the entire debris reservoir (i.e. $W_{\rm fr}/M_{\rm fr}$), while treating per-cluster values as approximate diagnostics.}
 The cluster mass is initially $m_c^{\rm raw}=\sum_{j\in c} w_j$. To mimic unresolved fine dust, we apply a controllable dust-removal factor to each cluster mass at this stage. Finally, we renormalise all $m_c$ so that $\sum_c m_c = M_{\rm fr}$, and scale cluster water fractions to preserve the total debris water mass: defining $S=\sum_c m_c w_{{\rm f},c}$ and the target $W_{\rm fr}$ from Sect.~\ref{sec:sph-table}, we set $w_{{\rm f},c}\leftarrow (W_{\rm fr}/S)\,w_{{\rm f},c}$. This two-step scaling ensures exact conservation of both total mass and water mass across survivors and debris. {In this way, differences in SPH particle resolution (and thus in the raw number/mass of catalogue entries) primarily affect the sampling of debris phase space, while the realised N-body debris masses are set by the interpolated budgets and the explicit resolution control $M_{\rm FM}$ (and any chosen dust-removal factor).}

\subsection{Frame mapping: from SPH to the simulation SoC}
\label{sec:frames}

The SPH outputs are given relative to the collision centre-of-mass (CoM) in a local spherical frame. To embed outcomes consistently in the global simulation frame, we compute a deterministic rotation that aligns the SPH “$y$-axis’’ with the larger body's velocity vector a finite time prior to impact, following these steps:

\begin{enumerate}
\item Form the relative state $(\boldsymbol{r}_0,\boldsymbol{v}_0)$ at the instant of contact, expressed in the smaller body's frame, i.e., $\boldsymbol{r}_0=\boldsymbol{x}_{\rm large}-\boldsymbol{x}_{\rm small}$ and $\boldsymbol{v}_0=\boldsymbol{v}_{\rm large}-\boldsymbol{v}_{\rm small}$.
\item Rotate so that the specific angular momentum $\boldsymbol{h}=\boldsymbol{r}_0\times \boldsymbol{v}_0$ lies along $+\hat{\boldsymbol{z}}$; from $(\phi,\theta)$ of $\boldsymbol{h}$ derive the yaw/pitch that accomplishes this alignment.
\item Compute the Keplerian elements $(a,e,f_0)$ of the relative orbit in the aligned plane and rotate by $\omega$ such that $f=0$ occurs at $y=0$ (placing periapsis on the $x$-axis).
\item Back-trace along the conic to the distance $r_{\rm min}=5\,(R_1+R_2)$ (radii at impact) and take the unit vector of the large body's velocity there as the SPH $y$-axis proxy.
\item Map this proxy back into the original simulation frame to obtain two Euler angles $(\chi,\psi)$ that rotate SPH outputs into the simulation frame; if $h_z>0$ an extra 180$^{\circ}$ rotation about $\hat{\boldsymbol{y}}$ is applied to ensure a right-handed mapping.
\end{enumerate}

We then convert SPH spherical coordinates $(r,\theta,\phi)$ to Cartesian, rotate by $(\chi,\psi)$, and translate by the instantaneous collision CoM state $(\boldsymbol{x}_{\rm CoM},\boldsymbol{v}_{\rm CoM})$ before updating particles. For the radius--mass relation we use $R=C (m/M_{\oplus})^{S}$ with $C=1.008\,R_{\oplus}$ and $S=0.279$; dependence on water fraction is presently neglected at the radii level although these effects should be negligible.

\subsection{Reaccretion checks}
\label{sec:reaccretion}

Immediately after instantiating survivors and debris in the global frame we iterate simple pairwise reaccretion tests:

\begin{enumerate}
\item \emph{Survivor--survivor:} if two survivors have relative speed $v_{\rm rel}$ and separation $d$, and $v_{\rm rel}^2<2(m_1+m_2)/d$, we merge them inelastically by conserving linear momentum and water mass.
\item \emph{Debris--survivor:} for each debris piece, if $v_{\rm rel}^2<2(m_{\rm surv}+m_{\rm deb})/d$ with respect to any survivor, we accrete it onto that survivor. After any accretion, we re-test the survivor--survivor condition once.
\end{enumerate}

These checks provide a minimal correction for immediate bound re-collisions within the same time step, without introducing new integration sub-steps.

\subsection{Activation thresholds and test particles}

Debris are sorted by mass and assigned labels \texttt{D1}, \texttt{D2}, \dots. Immediately before appending debris, we set \texttt{N\_active} equal to the number of debris with $m\ge m_{\rm act}$ (default $m_{\rm act}=3\times10^{-7}\,M_{\odot}\approx0.1\,M_{\oplus}$). Particles with index $\ge\texttt{N\_active}$ are treated by \texttt{Rebound} as test particles (\texttt{sim.testparticle\_type=1}): they feel the gravity of active bodies but not of each other, which stabilises performance while preserving the gross dynamical influence of the ejecta cloud. Survivors are not added to \texttt{N\_active} by a separate increment; because they occupy earlier indices and debris are appended afterward, survivors typically remain within the active range, following index ordering rules in \texttt{Rebound} rather than an explicit survivor-specific rule. This activation policy, together with $M_{\rm FM}$, sets the effective mass--space resolution of the collisional cascade.

\subsection{Special cases and bookkeeping}
\label{sec:special}

Impacts involving the Sun or designated gas giants (indexed among the first $N_{\rm GG}$ bodies) are treated as perfect mergers: the secondary is merged into the primary, and the event is logged. Impacts with test-particle debris default to perfect merging with the active body. All collision events append a line to a structured table recording: time; $(v_0,\alpha,M_{\rm tot},\gamma,w_{\rm f,t},w_{\rm f,p})$; CoM state; rotation angles $(\chi,\psi)$ and $h_z$; survivor count and properties; integrated debris mass and water fraction; and the list of labels assigned to the newly created debris. Additionally, a full system snapshot (including Cartesian states) is written at each collision time, enabling the exact restart of a run from any event.

\subsection{Numerical safety and diagnostics}

To guard against spurious contact detections, the resolver first rejects pairs with instantaneous separation $>\,2(R_1+R_2)$ (where $R_i$ is computed from the current $m_i$). Energy is tracked with \texttt{sim.track\_energy\_offset=1}. A progress log records wall-clock timing and the relative energy error at each output. Optional 2D/3D visual diagnostics of the pre- and post-impact geometry can be produced for debugging.

\section{Test case}
\subsection{Our code vs SyMBA}
\label{symba}
\begin{figure}[h]
    \centering
    \includegraphics[width=0.9\linewidth]{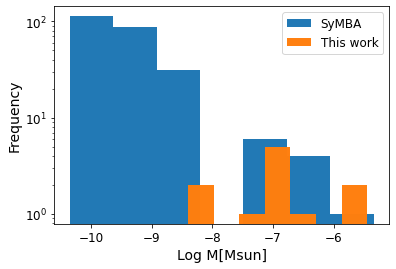} 
    \caption{Body mass distributions at the end of the two simulations. The \texttt{SyMBA} simulation produces more low-mass debris, whereas our simulation yields a more top-heavy distribution.}
    \label{fig:example1}
\end{figure}

\begin{figure}[h]
    \centering
    \includegraphics[width=0.9\linewidth]{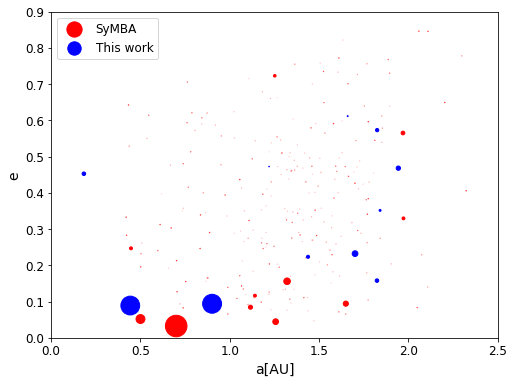} 
    \caption{Semimajor axis–eccentricity distributions of the bodies at the end of the simulations. Point sizes are scaled by object mass. In both cases, the distributions become dynamically hotter at larger semimajor axes. Our simulation also exhibits a more top-heavy mass distribution than the \texttt{SyMBA} simulation.}
    \label{fig:example2}
\end{figure}

We test our code through a comparison against \cite{morby} who used  \texttt{SyMBA} \citep{symba} to study the formation of Mercury from excited initial conditions.
We hence setup a simulation equivalent to their nominal case of a flat annulus planetesimals disk (between 0.7 to 1 AU) containing 160, 0.025 M$_\oplus$  embryos, with an average eccentricity of 0.1 drawn from a Rayleigh distribution.  \\
As a first and basic sanity check, we compare the total mass remaining at the end of the simulations (at 100 Myr) and we find M$_{tot,SyMBA}$/M$_{tot,here}$=0.88, indicating a small difference of only 12\%. \\ For an in depth comparison of the objects mass distributions we show the masses histogram in Fig. \ref{fig:example1}. This plot shows a slightly more top heavy distribution and a lack in leftover small planetesimals for our code compared to \texttt{SyMBA}. This is not surprising since a central feature of our approach is the clustering of small debris into bigger objects. Additionally, in Fig. \ref{fig:example2} we show the final semimajor axis vs eccentricity of all objects. While, qualitatively, the two distributions seem to share the same general features (the presence of two dominant objects in addition to a tail of smaller hotter objects at larger semimajor axis), our distribution seem to be dynamically hotter overall. 
{ We interpret these offsets as a consequence of the debris-resolution prescription rather than merely a stochastic difference between runs. In \texttt{SHARD}, the SPH-interpolated debris mass and water budgets are conserved, but the unresolved reservoir is represented by \(N_{\rm fr}=\lfloor M_{\rm fr}/M_{\rm FM}\rfloor\) mass-weighted clusters; thus the choice of \(M_{\rm FM}\) controls how the same debris is dynamically discretized. Concentrating debris into fewer, more massive tracers makes individual clumps easier to reaccrete through the local energy criterion and leaves less long-lived low-mass material to provide dynamical damping. This naturally explains the more top-heavy mass spectrum, the deficit of leftover planetesimals, part of the higher retained mass relative to debris-loss prescriptions such as \texttt{SyMBA}, and the dynamically hotter final state. It also implies that fragment resolution is physically consequential for Mercury formation, where the survival, mass, and core-mass fraction of a Mercury analogue depend on whether stripped mantle-rich debris is reaccreted or removed; by the same mechanism, water-rich ejecta can either be returned to growing planets or lost from the resolved system.}

Finally, we implement the same ``nine collisional types'' postprocessing code to track and evolve the core-mantle composition of the objects. Our results are found to be consistent with \cite{morby} with Mercury-analogues forming with the right mass and orbit starting from an Earth-like core mass fraction.   

\subsection{Water evolution}
\begin{figure}[h]
    \centering
    \includegraphics[width=0.7\linewidth]{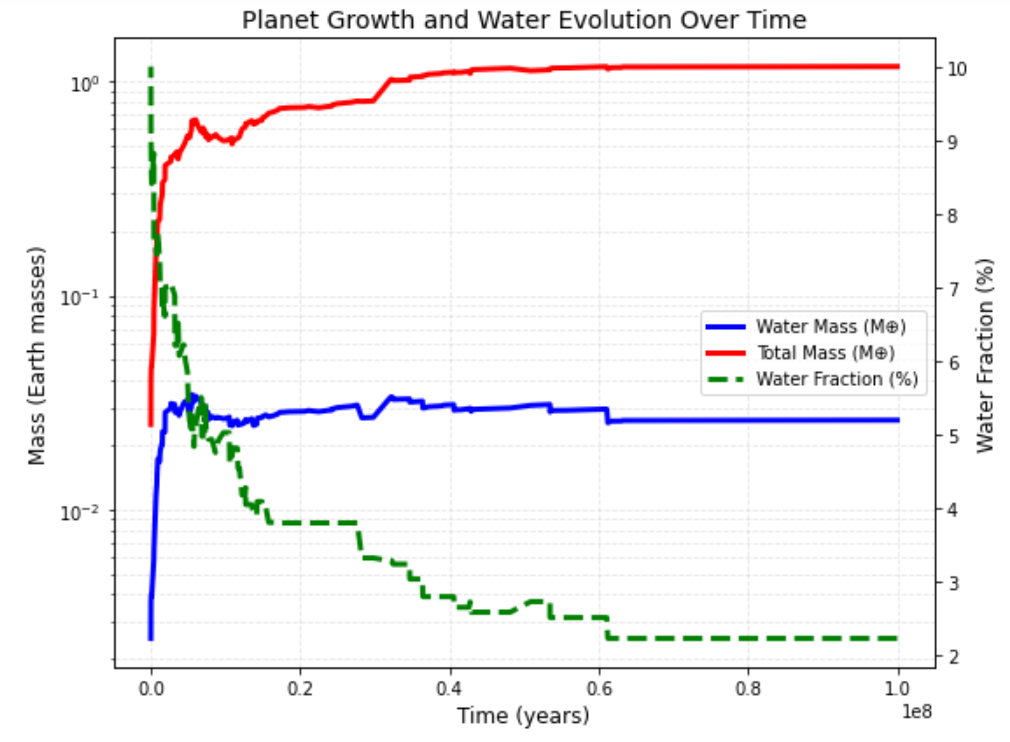} 
    \caption{The time evolution of the water and total masses for our Earth analogue. While both total and water masses increase with time, the water mass fraction decreases monotonously due to planetesimals losing water-rich debris. }
    \label{water1}
\end{figure}

{
In this section we study the water mass evolution in our simulation through the example of one of the Earth-like objects we obtained. The planet (object E125) is 1.17 M$_\oplus$ at 0.44 AU. The evolution of the planet's total and water masses, in addition to the water mass fraction, as shown in Fig. \ref{water1}. This figure shows that while the 2 masses are increasing as a function of time, the water mass fraction is decreasing. To understand this behavior we analyze some of the events and collisions that created and shaped the planet. In Table \ref{coll1} we show the first collision that E125 was involved in. E125's mass increased by a factor of 65\%, while its water mass fraction decreased by 10\%. This is due to losing water-rich debris, where we find that the average debris has a water mass fraction of 65.89\%.  

}

\begin{table}[h!]
\centering
\begin{tabular}{lccc}
\hline
{Particle} & {Mass ($M_{\oplus}$)} & {Water Fraction (\%)} & {Stage} \\
\hline
E125 & 0.024974 & 10.000 & Involved \\
E89  & 0.024974 & 10.000 & Involved \\
\hline
E125 & 0.041368 & 9.098 & Resulting \\
E89  & 0.008099 & 9.570 & Resulting \\
\hline
\end{tabular}
\caption{Collision event at $t = 9720.38$ yr. Particles E125 and E89 of equal mass and water fraction collide resulting in mass transfer and a decrease in water fraction for both due to losing water rich debris.}
\label{coll1}
\end{table}

\section{Summary \& conclusions}
We have introduced and validated a collision–resolution framework for late-stage terrestrial planet formation that (i) maps detected $N$-body impacts to a six‑parameter catalogue of smoothed‑particle hydrodynamics (SPH) outcomes and (ii) statistically reconstructs the otherwise unresolved debris into a small number of dynamically consistent fragments by mass‑weighted $k$‑means clustering in velocity space. The method is implemented on top of \textsc{REBOUND}/\texttt{mercurius}, and the code is available for use at \href{https://github.com/malidib/shard}{https://github.com/malidib/shard} .

\subsection{Summary of the collision-resolution pipeline}
\label{sec:pipeline-summary}

Given a detected contact between two non-test bodies, the pipeline is:

\begin{enumerate}
\item Compute $(v_0,\alpha,M_{\rm tot},\gamma,w_{\rm f,t},w_{\rm f,p})$ and the collision CoM state.
\item Interpolate within the SPH table to obtain the two largest remnants (mass and water fractions, local SPH-frame kinematics), drop sub-resolution survivors, and enforce water/mass clipping.
\item Compute the debris mass budget $M_{\rm fr}$ and water budget $W_{\rm fr}$.
\item Interpolate SPH debris catalogues at the same query point to assemble a weighted debris sample.
\item  In velocity space, cluster into $N_{\rm fr}$ fragments with mass-weighted centroids; apply the dust factor and renormalise to match $M_{\rm fr}$ and $W_{\rm fr}$ exactly.
\item Map survivors and debris from the SPH frame to the simulation frame via $(\chi,\psi)$, then add them to the integrator (survivors replacing the colliders; debris appended).
\item Run instantaneous survivor--survivor and debris--survivor reaccretion checks.
\item Set the activation split based on $m_{\rm act}$; label debris \texttt{D1--D$N_{\rm fr}$}.
\item Log the event and write a collision-time snapshot.
\end{enumerate}

This strategy ensures that every impact outcome is dynamically consistent with nearby SPH simulations while maintaining tractable particle counts and strict conservation of mass and volatile inventories, modulo the deliberate dust-removal parametrisation.

\subsubsection{Reproducibility knobs and defaults}

The main numerical and physical controls are:
\begin{itemize}
\item Integration: $\Delta t=0.01\,$yr, $h_{\mathrm{fac}}=5$, $\mathrm{min\_dt}=10^{-4}\Delta t$, collision geometry \texttt{direct}, boundary \texttt{none}.
\item Fragment resolution: $M_{\rm FM}=5.5\times10^{-4}\,M_{\oplus}$; activation threshold $m_{\rm act}=3\times10^{-7}\,M_{\odot}$.
\item Debris clustering: $N_{\rm fr}=\lfloor M_{\rm fr}/M_{\rm FM}\rfloor$; \texttt{KMeans} in velocity space with sample weights $=\,$debris mass; post-clustering dust factor $=1/2$.
\item Removal: unbound orbits; $a>15$\,AU; $a<0.1$\,AU or $q<0.01$\,AU.
\end{itemize}

\subsection{Limitations \& future work}
Multiple modelling choices delimit the present scope and can be improved in future works:
\begin{itemize}
    \item clustering is performed solely in velocity space, capturing the dominant dynamical structure but not fine spatial substructure within the ejecta.
    \item the immediate reaccretion tests are purely energetic and local to the collision timestep, leaving longer‑term gravitational focusing to the integrator. 
    \item the radius–mass relation neglects composition and water‑fraction dependencies at fixed mass.
    \item debris interpolation is restricted to the convex hull of the SPH grid; behaviour at more extreme impact parameters awaits expanded SPH coverage.
    \item the adopted dust‑removal factor and resolution thresholds ($M_{\rm FM}$, activation split) are explicit, tunable parameters that trade physical fidelity against computational cost.
\end{itemize}

\section*{Acknowledgements}
We thank the anonymous referee for their useful comments that helped improve the clarity and quality of this manuscript. We thank A. Morbidelli, D. Valencia, and J. Scora for insightful discussions on Mercury's formation, as well as for their support with code comparison.
This material is based on work supported by Tamkeen under the NYU Abu Dhabi Research Institute grant CASS.

\newpage
\appendix

\section{Methods: SPH data}
\label{sec:data}

This section documents the \emph{SPH data used by the code} and how it is prepared for consumption by the collision resolver. The implementation relies exclusively on a library of smoothed‑particle hydrodynamics (SPH) collision outcomes sampled on a discrete grid in six control parameters, plus per‑collision debris catalogues. 

\subsection{SPH outcome library: scope and provenance}
\label{sec:data:provenance}

Each datum corresponds to a single SPH impact between two non‑rotating bodies and is indexed by the six‑tuple
\[
\theta \equiv \bigl(v_0,\ \alpha,\ M_{\rm tot},\ \gamma,\ w_{f,t},\ w_{f,p}\bigr),
\]
where $v_0\equiv v_{\rm imp}/v_{\rm esc}$ is the impact speed normalized by mutual escape speed, $\alpha$ is the angle between the relative position and velocity vectors at contact (head‑on $\approx0^\circ$, grazing $\approx90^\circ$), $M_{\rm tot}$ is the total colliding mass, \\ $\gamma\!=\!\min(m_1/m_2,\ m_2/m_1)$, and $w_{f,t}$, $w_{f,p}$ are the bulk water mass fractions (\%) of target and projectile. The library samples these parameters on a fixed grid (Table~\ref{tab:grid}). 

{SPH snapshots are recorded a short time after impact (order $\sim$10~hr in the source simulations) and then reduced into the tabular formats described below; the code treats these as fixed inputs and does not depend on the hydrodynamic solver internals. The adopted library is homogeneous in resolution (of order $\sim2\times10^4$ SPH particles per collision), but this resolution remains modest: convergence is expected to be poorest for the water partitioning of the second-largest remnant and for the smallest clumps in the post-processed debris. Accordingly, we treat remnant- and debris-level water fractions as approximate (potentially uncertain at the $\gtrsim$\,tens-of-percent level for the second remnant, and larger for smaller fragments), and we focus on conserving the {\em total} post-impact water mass budget when mapping outcomes into the $N$-body code (Sect.~\ref{sec:kmeans}). In particular, the most robust debris quantity is the mass-weighted mean water fraction of the debris reservoir, whereas per-fragment (cluster) water fractions should be interpreted as bookkeeping diagnostics rather than precise compositional predictions.}

\begin{table}[h]
\centering
\caption{Discrete grid used by the SPH outcome library and its on‑disk code encoding.}
\label{tab:grid}
\begin{tabular}{l l l}
\hline
Parameter & Grid values & Code digit mapping \\
\hline
$v_0$ & $\{1,\ 1.5,\ 2,\ 3,\ 5\}$ & 0$\!\to\!$1, 1$\!\to\!$1.5, 2$\!\to\!$2, 3$\!\to\!$3, 4$\!\to\!$5 \\
$\alpha$ (deg) & $\{0,\ 20,\ 40,\ 60\}$ & 0$\!\to\!$0, 1$\!\to\!$20, 2$\!\to\!$40, 3$\!\to\!$60 \\
$M_{\rm tot}$ & $\{2M_{\rm Ceres},\ 2M_{\rm Moon},\ 2M_{\rm Mars},\ 2M_{\oplus}\}$ & 0$\!\to\!$2$M_{\rm Ceres}$, 1$\!\to\!$2$M_{\rm Moon}$, 2$\!\to\!$2$M_{\rm Mars}$, 3$\!\to\!$2$M_{\oplus}$ \\
$\gamma$ & $\{0,\ 0.5,\ 1\}$ & 0$\!\to\!$0, 1$\!\to\!$0.5, 2$\!\to\!$1 \\
$w_{f,t}$ (\%) & $\{10,\ 20\}$ & 0$\!\to\!$10, 1$\!\to\!$20 \\
$w_{f,p}$ (\%) & $\{10,\ 20\}$ & 0$\!\to\!$10, 1$\!\to\!$20 \\
\hline
\end{tabular}

\vspace{3pt}
\small The six digits are concatenated into a base ``544322'' code (bases for $v_0/\alpha/M_{\rm tot}/\gamma/w_{f,t}/w_{f,p}$ respectively), used throughout the on‑disk filenames and table rows.
\end{table}

\subsection{On‑disk layout and column semantics}
\label{sec:data:files}

All inputs live alongside the integrator in two artefacts:

\paragraph{\texttt{SPH.table} (one line per SPH collision).} Each row contains:
\begin{enumerate}
\item an integer \texttt{id};
\item a six‑digit \texttt{code} as in Table~\ref{tab:grid};
\item the six real‑valued parameters $(v_0,\alpha,M_{\rm tot},\gamma,w_{f,t},w_{f,p})$;
\item \texttt{Nbig}: nominal number of surviving macroscopic bodies (special value $-1$ flags catastrophic disruption, ``crashed'');
\item $m_{\rm fr}$: total fragmented mass fraction relative to $M_{\rm tot}$;
\item two \emph{largest} survivors, each stored as:
\begin{itemize}
\item position in the SPH centre‑of‑mass (CoM) frame, spherical coordinates $[r,\ \theta,\ \phi]$ with $r$ in AU and angles in rad,
\item velocity in the SPH CoM frame, spherical components $[v,\ \theta_v,\ \phi_v]$ with $v$ in AU/yr$/2\pi$ and angles in rad,
\item $m_i$: survivor mass fraction relative to $M_{\rm tot}$,
\item $W_i$: survivor water mass fraction relative to the total pre‑impact water mass.
\end{itemize}
The second survivor may be marked absent by $m_2=-1$ (Perfect Merging, PM).
\end{enumerate}

\paragraph{\texttt{SPHDebris\_catalogue/\{id\}\_\{code\}.dat} (per‑collision debris list).}
A plain‑text table with columns
\[
(x,\ y,\ z,\ v_x,\ v_y,\ v_z,\ m,\ m/M_{\rm tot},\ w_f),
\]
all in REBOUND units (AU, AU/yr$/2\pi$, dimensionless fractions). The first \texttt{Nbig} rows correspond to the survivors for that SPH collision; the rows thereafter are debris fragments. {These debris rows comprise the complete SPH fragment list beyond the survivors down to the smallest entries present in the SPH output (no additional ``resolved-body'' selection is applied at the catalogue stage).} When present, this file is loaded lazily on demand.

\subsection{Units and conventions}
\label{sec:data:units}

All quantities are in REBOUND units with $G=1$, distance in AU, mass in $M_\odot$, and time in yr$/2\pi$. Earth mass is $M_{\oplus}\equiv{\rm MEAR}=3.003\times10^{-6}M_\odot$. Positions/velocities in \texttt{SPH.table} are expressed in the local SPH CoM frame; debris files provide Cartesian $(\mathbf{x},\mathbf{v})$ in the same frame. Water fractions in \texttt{SPH.table} are normalized to the total pre‑impact water mass; debris \texttt{wf} is per‑particle bulk water fraction in percent.

\subsection{Indexing and code lookup}
\label{sec:data:index}

The code maps the six‑digit \texttt{code} to a unique integer index via a mixed‑radix scheme (``544322'') and uses it both to sanity‑check corner selections during interpolation and to locate the matching debris file \texttt{\{id\}\_\{code\}.dat}.

\subsection{Preparation for consumption: interpolation payloads}
\label{sec:data:prep}

At run time, the resolver queries the SPH library at an arbitrary $\theta$ extracted from the instantaneous pre‑impact N‑body state.

\paragraph{Two largest remnants (from \texttt{SPH.table}).}
The code gathers the $2^6$ neighbouring grid corners in parameter space and performs chained binary interpolations to return the two largest bodies in the SPH CoM frame:
\begin{itemize}
\item linear weights along $v_0,\ \alpha,\ \gamma,\ w_{f,t},\ w_{f,p}$;
\item \emph{logarithmic} weights along $M_{\rm tot}$;
\item angles are interpolated on the circle via
\[
\operatorname{interp}_\angle(a,b;d)=\operatorname{atan2}\bigl((1-d)\sin a+d\sin b,\ (1-d)\cos a+d\cos b\bigr);
\]
 \item special handling keeps PM states consistent, replaces invalid negative masses by small positive fallbacks ($1/\texttt{SPHRES}$) and, for PM, may set $m_1=1$; it enforces $\sum_i m_i\le 1$, zeros invalid magnitudes in $r$ and $v$, and renormalizes water fractions (setting $W_1=1$ for PM and otherwise clipping $W_i\ge 0$ and rescaling so that $W_1+W_2\le 1$).
\end{itemize}
{This yields $\{(\mathbf{x}_i^{\rm sph},\mathbf{v}_i^{\rm sph},\,m_i/M_{\rm tot},\,W_i/W_{\rm tot})\}_{i=1,2}$ ready for unit conversion and frame mapping, where $\mathbf{x}_i^{\rm sph}$ and $\mathbf{v}_i^{\rm sph}$ are the SPH-frame position/velocity vectors (stored in spherical components in \texttt{SPH.table}).}

\paragraph{Small‑fragment phase space (from debris catalogues).}
To approximate the local debris distribution at $\theta$, the code \emph{combines} debris from the same $2^6$ corners using the product of the 1D interpolation weights, but \emph{without} extrapolation beyond the tabulated grid; SPH collisions flagged as catastrophic (``crashed'') are skipped {in the phase-space sampling because their debris catalogues are unavailable or unreliable, while their influence is still carried by the interpolated scalar debris budgets ($m_{\rm fr}$ and, where applicable, $W_{\rm fr}$)}. From each contributing debris file it discards the first \texttt{Nbig} rows (those are the two survivors already provided by \texttt{SPH.table}) and retains the rest as weighted samples of $(\mathbf{x},\mathbf{v})$ with mass weights {$\Lambda\cdot(m/M_{\rm tot})$} and per‑particle water fractions $w_f$. The merged list is persisted to a human‑readable \texttt{.debris} diagnostic file bundled with the simulation outputs.

\section{Diagnostic tests}
\subsection{Numerical diagnostics}
In this section we show the results for a series of numerical diagnostic tests we ran for the same example shown in section \ref{symba}.\\

\begin{figure}[h]
    \centering
    \includegraphics[width=0.7\linewidth]{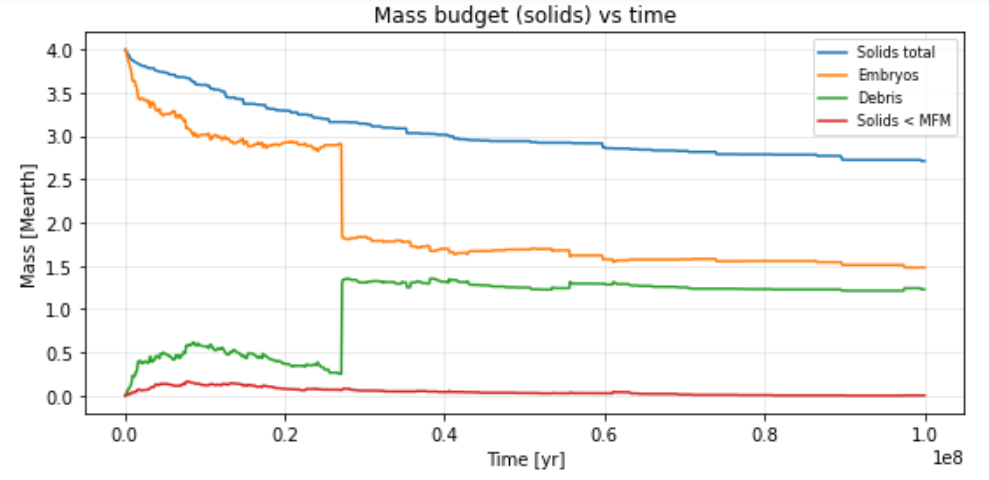} 
    \caption{Evolution of the solid mass budget over time in our nominal simulation. A catastrophic collision between embryos occurs around 25 Myr, causing a sharp decrease in embryo mass that is offset by a corresponding increase in debris mass.}
    \label{fig:diag1}
\end{figure}

\begin{figure}[h]
    \centering
    \includegraphics[width=0.7\linewidth]{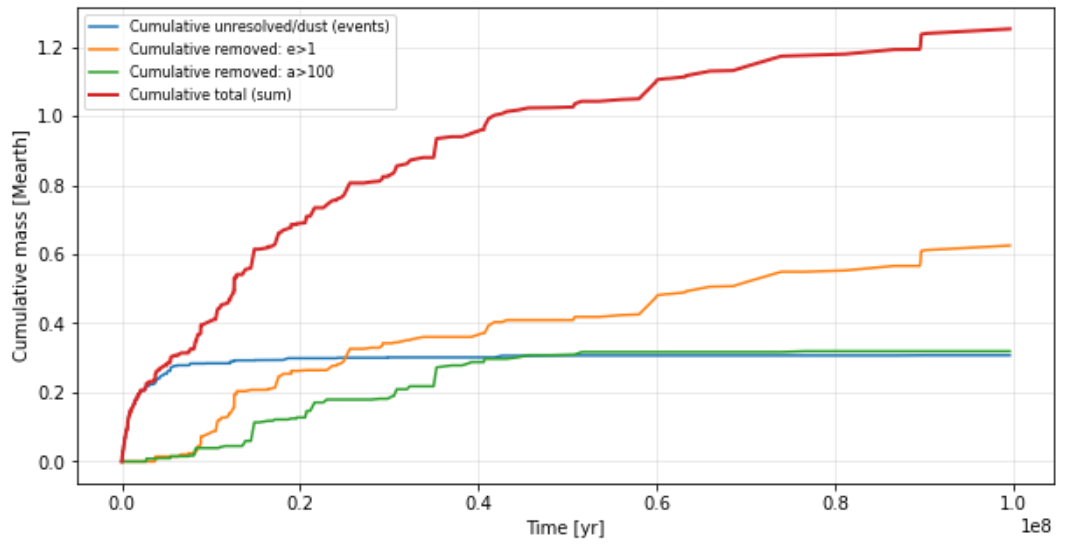} 
    \caption{The cumulative mass removed from the simulation over time, separated by sink type. Mass loss as unresolved dust dominates initially, but later plateaus and is overtaken by mass ejected hyperbolically from the solar system.}
    \label{fig:diag8}
\end{figure}

\begin{figure}[h]
    \centering
    \includegraphics[width=0.7\linewidth]{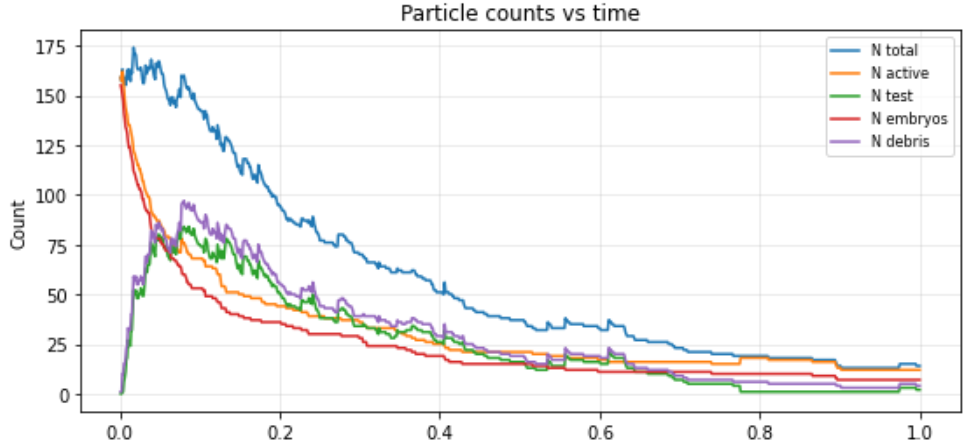} 
    \caption{Time evolution of the number of particles in the simulation, by type. While all particles start as equal mass active embryos, these get replaced by test debris as the simulation progresses. }
    \label{fig:diag2}
\end{figure}

In Figs. \ref{fig:diag1}, \ref{fig:diag8}, and \ref{fig:diag2} we show the time evolution of the solid mass budget and particles number for our nominal simulation.

In our code, the resolved solids inventory decreases through (i) dynamical excision of particles with \(e\ge 1\) or \(a>15\,\mathrm{AU}\), (ii) boundary-driven mergers into the Sun (and perfect mergers with gas giants) for \(a<0.1\,\mathrm{AU}\) or \(q<0.01\,\mathrm{AU}\), and (iii) unresolved fragmentation when \(M_{\rm fr}<M_{\rm FM}\) (so \(N_{\rm fr}=0\)). Therefore, as seen in Figs. \ref{fig:diag1}, \ref{fig:diag8}, and \ref{fig:diag2}, the total solids mass is conserved as the missing mass in Fig. \ref{fig:diag1} is either transferred to the Sun/giants, removed as ejected/out-of-bounds material, or relegated to an untracked sub-resolution dust component.

In Fig. \ref{fig:diag1}, at t $\sim$ 25 Myr, we see a significant dip in the Embryos mass, compensated for by an increase in debris mass. 

\begin{figure}[h]
    \centering
    \includegraphics[width=0.7\linewidth]{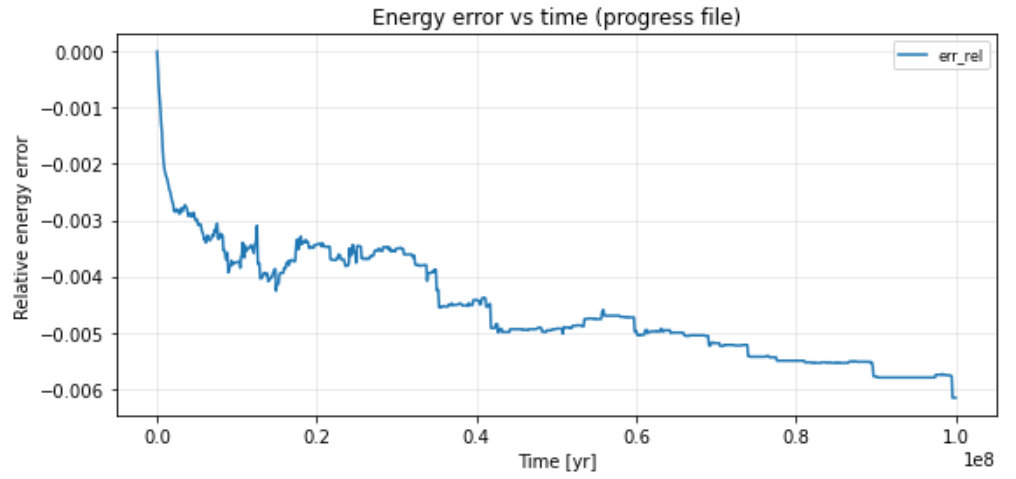} 
    \caption{The time evolution of the nominal simulation's relative energy error. }
    \label{fig:diag3}
\end{figure}

In Fig. \ref{fig:diag3} we show the time evolution of the energy error relative to that at the start of the simulation ($(E_t-E_0)/E_0$). Ideally, for a perfectly energy-conserving integrator, this value would remain constant throughout the simulation. A significant change indicates a systematic energy drift, which may arise from numerical integration errors, finite time-stepping, close encounters, or the treatment of collisions and mergers. Monitoring this relative error provides a quantitative measure of the long-term stability and accuracy of the N-body integration scheme. In our case, the relative error exhibits negligible drift in the order of $\sim 10^{-3}$.

\begin{figure}[h]
    \centering
    \includegraphics[width=0.7\linewidth]{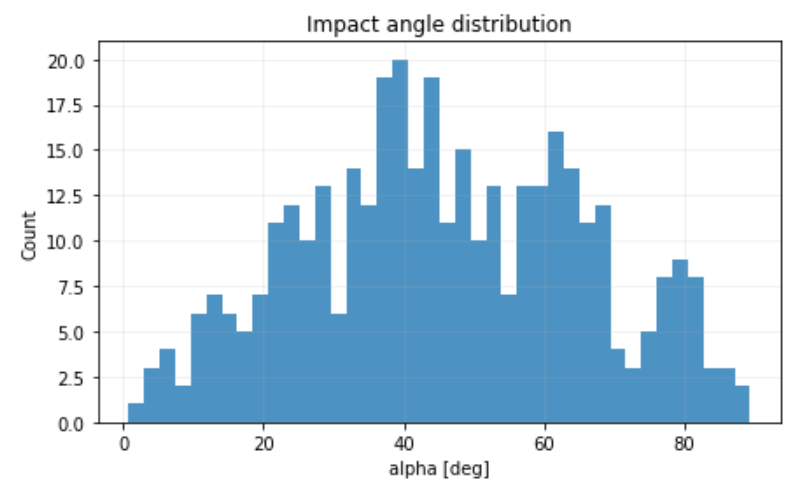} 
    \caption{Distribution of all impact angles in our nominal simulation. The average is 45$^o$ as expected.}
    \label{fig:diag4}
\end{figure}

In Fig. \ref{fig:diag4} we show the collisions impact angle distribution for the same simulation. As expected, this follows a near gaussian distribution with a mean of 45$^o$.

\subsection{SPH parameter space}
\begin{figure}[h]
    \centering
    \includegraphics[width=0.7\linewidth]{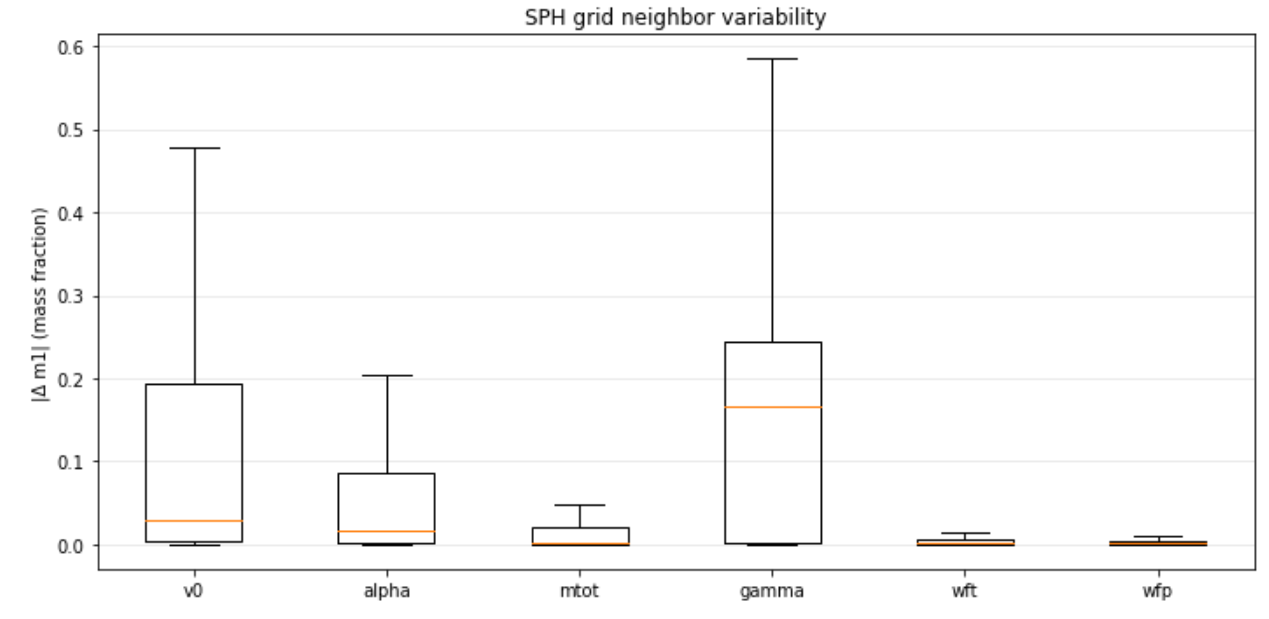} 
    \caption{The effect a variety of variables have on the mass of the largest remnant post collision.}
    \label{fig:diag9}
\end{figure}

\begin{figure}[h]
    \centering
    \includegraphics[width=0.7\linewidth]{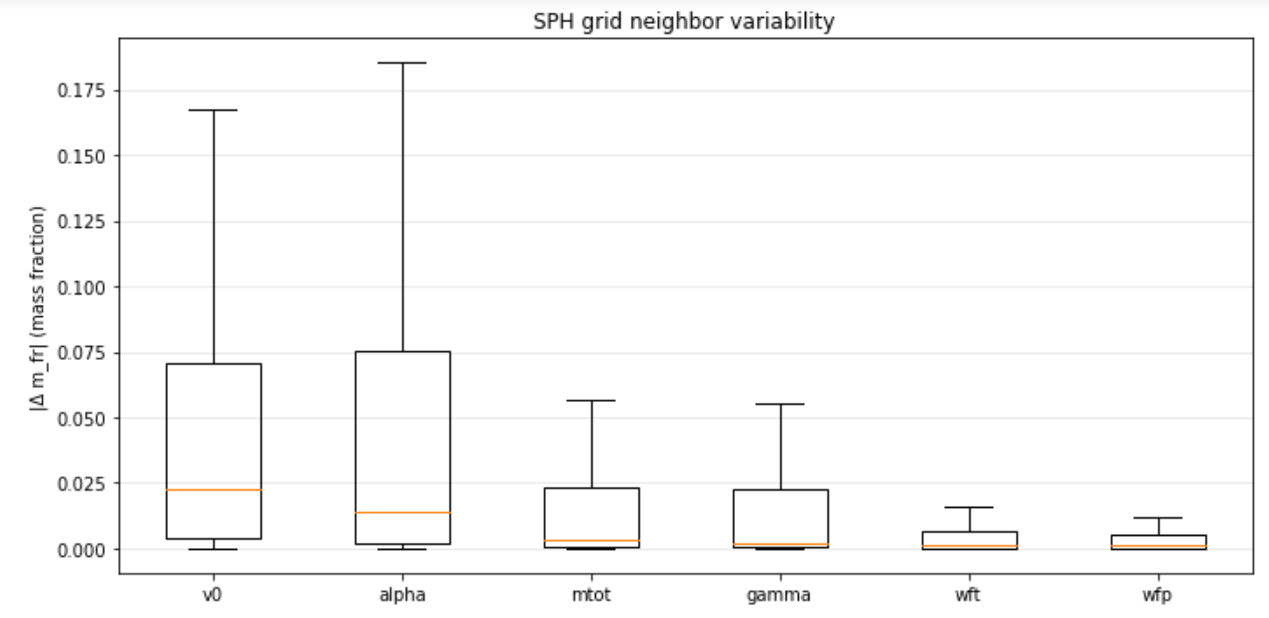} 
    \caption{The effect a variety of variables have on the mass of the fragments post collision.}
    \label{fig:diag10}
\end{figure}

{In Figs. \ref{fig:diag9} and \ref{fig:diag10} we quantified local grid variability for the two outputs ($m_{\rm fr}$ and $M_1$) by taking absolute differences between immediate neighbors along each grid axis and summarizing those differences with the 16th percentile (p16), median, 84th percentile (p84), and maximum.} For $m_{\rm fr}$, typical neighbor-to-neighbor changes are small (medians $\sim10^{-3}$--$3\times10^{-3}$ for \texttt{wfp}, \texttt{wft}, \texttt{gamma}, and \texttt{mtot}), increase for \texttt{alpha} and \texttt{v0} (medians $0.0139$ and $0.0228$), and nevertheless exhibit a pronounced upper tail (e.g., p84 up to $0.166$ and maxima $\simeq0.82$--$0.91$). For $M_1$, composition steps remain modest (\texttt{wft}/\texttt{wfp} medians $1.83\times10^{-3}$/$1.15\times10^{-3}$), while dynamical axes drive much larger typical variability, most notably \texttt{gamma} (median $0.167$, p84 $0.257$) and \texttt{v0} (median $0.0289$, p84 $0.336$), with maxima approaching $0.86$. The consistent gap between very small p16 values (often $\lesssim10^{-3}$) and much larger p84 values ($\gtrsim0.1$ in several cases) shows that most neighbor steps are gentle, but a minority of adjacent grid cells produce sharp local jumps that dominate the extrema. These seem to correspond to some one-step perturbations crossing boundaries between collisional outcome classes.

\bibliographystyle{aasjournal}
\bibliography{refs}

@ARTICLE{symba,
       author = {{Duncan}, Martin J. and {Levison}, Harold F. and {Lee}, Man Hoi},
        title = "{A Multiple Time Step Symplectic Algorithm for Integrating Close Encounters}",
      journal = {\aj},
         year = 1998,
        month = oct,
       volume = {116},
       number = {4},
        pages = {2067-2077},
          doi = {10.1086/300541},
       adsurl = {https://ui.adsabs.harvard.edu/abs/1998AJ....116.2067D}
}

@ARTICLE{rebound,
       author = {{Rein}, H. and {Liu}, S.-F.},
        title = "{REBOUND: an open-source multi-purpose N-body code for collisional dynamics}",
      journal = {\aap},
         year = 2012,
        month = jan,
       volume = {537},
          eid = {A128},
        pages = {A128},
          doi = {10.1051/0004-6361/201118085},
archivePrefix = {arXiv},
       eprint = {1110.4876},
 primaryClass = {astro-ph.EP},
       adsurl = {https://ui.adsabs.harvard.edu/abs/2012A&A...537A.128R}
}

@ARTICLE{mercurius,
       author = {{Rein}, Hanno and {Hernandez}, David M. and {Tamayo}, Daniel and {Brown}, Garett and {Eckels}, Emily and {Holmes}, Emma and {Lau}, Michelle and {Leblanc}, R{\'e}jean and {Silburt}, Ari},
        title = "{Hybrid symplectic integrators for planetary dynamics}",
      journal = {\mnras},
         year = 2019,
        month = jun,
       volume = {485},
       number = {4},
        pages = {5490-5497},
          doi = {10.1093/mnras/stz769},
archivePrefix = {arXiv},
       eprint = {1903.04972},
 primaryClass = {astro-ph.EP},
       adsurl = {https://ui.adsabs.harvard.edu/abs/2019MNRAS.485.5490R}
}

@article{Chambers2001,
  author  = {Chambers, John E.},
  title   = {Making More Terrestrial Planets},
  journal = {Icarus},
  year    = {2001},
  volume  = {152},
  number  = {2},
  pages   = {205--224},
  doi     = {10.1006/icar.2001.6639}
}

@article{OBrien2006,
  author  = {O'Brien, David P. and Morbidelli, Alessandro and Levison, Harold F.},
  title   = {Terrestrial Planet Formation with Strong Dynamical Friction},
  journal = {Icarus},
  year    = {2006},
  volume  = {184},
  number  = {1},
  pages   = {39--58},
  doi     = {10.1016/j.icarus.2006.04.005}
}

@article{AgnorAsphaug2004,
  author  = {Agnor, Craig B. and Asphaug, Erik},
  title   = {Accretion Efficiency During Planetary Collisions},
  journal = {The Astrophysical Journal Letters},
  year    = {2004},
  volume  = {613},
  pages   = {L157--L160},
  doi     = {10.1086/425158}
}

@article{LeinhardtStewart2012,
  author  = {Leinhardt, Zo{\"e} M. and Stewart, Sarah T.},
  title   = {Collisions Between Gravity-dominated Bodies. I. Outcome Regimes and Scaling Laws},
  journal = {The Astrophysical Journal},
  year    = {2012},
  volume  = {745},
  pages   = {79},
  doi     = {10.1088/0004-637X/745/1/79}
}

@article{Chambers2013,
  author       = {Chambers, John~E.},
  title        = {Late-stage planetary accretion including hit-and-run collisions and fragmentation},
  journal      = {Icarus},
  year         = {2013},
  volume       = {224},
  number       = {1},
  pages        = {43--56},
  doi          = {10.1016/j.icarus.2013.02.015},
  url          = {https://doi.org/10.1016/j.icarus.2013.02.015}
}

@article{Schafer2016,
  author       = {Sch{\"a}fer, Christoph~M. and Riecker, Sven and Maindl, Thomas~I. and Speith, Roland and Scherrer, Samuel and Kley, Wilhelm},
  title        = {A smooth particle hydrodynamics code to model collisions between solid, self-gravitating objects},
  journal      = {Astronomy \& Astrophysics},
  year         = {2016},
  volume       = {590},
  pages        = {A19},
  doi          = {10.1051/0004-6361/201528060},
  url          = {https://doi.org/10.1051/0004-6361/201528060}
}

@article{Leinhardt2012,
  author       = {Leinhardt, Zo{"e}~M. and Stewart, Sarah~T.},
  title        = {Collisions between gravity-dominated bodies. I. Outcome regimes and scaling laws},
  journal      = {The Astrophysical Journal},
  year         = {2012},
  volume       = {745},
  number       = {1},
  pages        = {79},
  doi          = {10.1088/0004-637X/745/1/79},
  url          = {https://doi.org/10.1088/0004-637X/745/1/79}
}

@article{Schafer2020,
  author       = {Sch{\"a}fer, Christoph~M. and Wandel, Oliver~J. and Burger, Christoph and Maindl, Thomas~I. and Malamud, Uri and Buruchenko, Sergey~K. and Sfair, Rafaela and Audiffren, Hugo and Vavilina, Ekaterina and Winter, Peter~M.},
  title        = {A versatile smoothed particle hydrodynamics code for graphic cards},
  journal      = {Astronomy and Computing},
  year         = {2020},
  volume       = {33},
  pages        = {100410},
  doi          = {10.1016/j.ascom.2020.100410},
  url          = {https://doi.org/10.1016/j.ascom.2020.100410}
}

@ARTICLE{reboundx,
       author = {{Tamayo}, Daniel and {Rein}, Hanno and {Shi}, Pengshuai and {Hernandez}, David M.},
        title = "{REBOUNDx: a library for adding conservative and dissipative forces to otherwise symplectic N-body integrations}",
      journal = {\mnras},
         year = 2020,
        month = jan,
       volume = {491},
       number = {2},
        pages = {2885-2901},
          doi = {10.1093/mnras/stz2870},
archivePrefix = {arXiv},
       eprint = {1908.05634},
 primaryClass = {astro-ph.EP},
       adsurl = {https://ui.adsabs.harvard.edu/abs/2020MNRAS.491.2885T}
}

@ARTICLE{Benz1994,
       author = {{Benz}, W. and {Asphaug}, E.},
        title = "{Impact Simulations with Fracture. I. Method and Tests}",
      journal = {Icarus},
         year = "1994",
        month = "Jan",
       volume = {107},
       number = {1},
        pages = {98-116},
          doi = {10.1006/icar.1994.1009},
       adsurl = {https://ui.adsabs.harvard.edu/abs/1994Icar..107...98B}
}

@article{Monaghan1985,
title = {Artificial viscosity for particle methods},
journal = {Applied Numerical Mathematics},
volume = {1},
number = {3},
pages = {187-194},
year = {1985},
issn = {0168-9274},
doi = {https://doi.org/10.1016/0168-9274(85)90015-7},
url = {https://www.sciencedirect.com/science/article/pii/0168927485900157},
author = {J.J. Monaghan and H. Pongracic}
}

@ARTICLE{morby,
       author = {{Scora}, Jennifer and {Valencia}, Diana and {Morbidelli}, Alessandro and {Jacobson}, Seth},
        title = "{Forming Mercury from Excited Initial Conditions}",
      journal = {\apj},
         year = 2024,
        month = may,
       volume = {967},
       number = {1},
          eid = {1},
        pages = {1},
          doi = {10.3847/1538-4357/ad39e6},
archivePrefix = {arXiv},
       eprint = {2404.17523},
 primaryClass = {astro-ph.EP},
       adsurl = {https://ui.adsabs.harvard.edu/abs/2024ApJ...967....1S}
}

@article{KokuboGenda2010,
  author  = {Kokubo, Eiichiro and Genda, Hidenori},
  title   = {Formation of Terrestrial Planets from Protoplanets Under a Realistic Accretion Condition},
  journal = {The Astrophysical Journal Letters},
  year    = {2010},
  volume  = {714},
  pages   = {L21--L25},
  doi     = {10.1088/2041-8205/714/1/L21}
}

@article{KobayashiDauphas2013,
  author  = {Kobayashi, Hiroshi and Dauphas, Nicolas},
  title   = {Small Planetesimals in a Massive Disk Formed Mars},
  journal = {Icarus},
  year    = {2013},
  volume  = {225},
  number  = {1},
  pages   = {122--130},
  doi     = {10.1016/j.icarus.2013.03.006}
}

@article{LevisonThommesDuncan2012,
  author  = {Levison, Harold F. and Duncan, Martin J. and Thommes, Edward},
  title   = {A Lagrangian Integrator for Planetary Accretion and Dynamics (LIPAD)},
  journal = {The Astronomical Journal},
  year    = {2012},
  volume  = {144},
  pages   = {119},
  doi     = {10.1088/0004-6256/144/4/119}
}

@article{MarcusStewartSasselovHernquist2009,
  author  = {Marcus, Robert A. and Stewart, Sarah T. and Sasselov, Dimitar and Hernquist, Lars},
  title   = {Collisional Stripping and Disruption of Super-Earths},
  journal = {The Astrophysical Journal Letters},
  year    = {2009},
  volume  = {700},
  number  = {2},
  pages   = {L118--L122},
  doi     = {10.1088/0004-637X/700/2/L118}
}

@article{StewartLeinhardt2012,
  author  = {Stewart, Sarah T. and Leinhardt, Zo{\"e} M.},
  title   = {Collisions Between Gravity-dominated Bodies. II. The Diversity of Impact Outcomes During the End Stage of Planet Formation},
  journal = {The Astrophysical Journal},
  year    = {2012},
  volume  = {751},
  pages   = {32},
  doi     = {10.1088/0004-637X/751/1/32}
}

@article{WalshLevison2019,
  author  = {Walsh, Kevin J. and Levison, Harold F.},
  title   = {Planetesimals to Terrestrial Planets: Collisional Evolution Amidst a Dissipating Gas Disk},
  journal = {Icarus},
  year    = {2019},
  volume  = {329},
  pages   = {88--100},
  doi     = {10.1016/j.icarus.2019.03.031}
}

@article{BurgerBazsoSchafer2020,
  author  = {Burger, Christoph and Bazs{\'o}, {\'A}kos and Sch{\"a}fer, Christoph M.},
  title   = {Realistic Collisional Water Transport During Terrestrial Planet Formation: Self-consistent Modeling by an N-body--SPH Hybrid Code},
  journal = {Astronomy \& Astrophysics},
  year    = {2020},
  volume  = {634},
  pages   = {A76},
  doi     = {10.1051/0004-6361/201936366}
}

\end{document}